\documentclass[RNAAS]{aastex631}

\begin{document}

\title{Revisiting the gas-phase chemical rate coefficients at high temperatures in CLOUDY}

\correspondingauthor{Gargi Shaw}
\email{gargishaw@gmail.com}

\author[0000-0003-4615-8009]{Gargi Shaw}
\affiliation{Department of Astronomy and Astrophysics\\ Tata Institute of Fundamental Research\\  
Mumbai 400005, India}

\author[0000-0003-4503-6333]{Gary Ferland}
\affiliation{Department of Physics and Astronomy\\ University of Kentucky\\
Lexington, KY 40506, USA}

\author[0000-0002-8823-0606]{M. Chatzikos}
\affiliation{Department of Physics and Astronomy\\ University of Kentucky\\
Lexington, KY 40506, USA}

\begin{abstract}
A two-body gas-phase reaction rate coefficient can be given by the usual Arrhenius-type formula 
which depends 
on temperature. 
The UMIST Database for Astrochemistry is a widely used database for reaction rate coefficients. 
They provide fittings for coefficients valid over a particular range of temperatures.
The permissible upper-temperature
limits vary over a wide range: from 100 K to 41000K. 
A wide range of temperatures occurs in nature; thus, it requires
evaluating the rate coefficients at temperatures outside the range of validity.
As a result, a simple extrapolation of the rate coefficients can lead to unphysically large values at high temperatures.
These result in unrealistic predictions. 
Here we present a solution to prevent the gas-phase reaction coefficients from 
diverging at a very high temperature. 

We implement this into the spectral synthesis code CLOUDY which operates over a wide range of temperatures from 
CMB to 10$^{10}$ K subject to different astrophysical environments.

\end{abstract}



\section{Introduction} \label{sec:intro}
In the UMIST Database for Astrochemistry (UDfA) \citep{{1997A&AS..121..139M},{2013A&A...550A..36M}}, 
a two-body gas-phase chemical reaction rate coefficient $k$ ($\rm{cm^3 \, s^{-1}}$) 
is given by the usual Arrhenius-type formula,
\begin {equation}
k=\alpha \left(\frac{T}{300}\right)^\beta \, \exp(-\gamma /T),
\end {equation}
where $T$ is the gas temperature. 
Rates for those reactions with $\gamma$ $<$ 0 will become unphysically large at low temperatures. 
Any species' predicted column densities and line intensities depend on the rate coefficients. 
Hence, it is important to use the right rate coefficients to predict the observables. 

\citet{2011A&A...530A...9R} has addressed the divergence  of rate coefficients at low temperatures. 
A similar problem occurs for high temperatures encountered with CLOUDY.
The spectroscopic simulation code CLOUDY simulates physical conditions in NLTE astrophysical
plasma using \textit{ab initio} microphysics over the entire electromagnetic range.  
Details about CLOUDY can be found in 
\citet{{2013RMxAA..49..137F},{2017RMxAA..53..385F},{2005ApJ...624..794S},{2005ApJS..161...65A},{2017ApJ...843..149S},
{2020RNAAS...4...78S},{2022ApJ...934...53S},
{2023RNAAS...7...45S}}
\footnote{\url{https://nublado.org}}.  
Rate coefficients with a positive $\beta$ can become large at high temperatures. 
That is the subject of this note.

The Leiden PDR workshop \citep{2007A&A...467..187R} compared  PDR codes and
chemical networks. 
The Leiden benchmark chemical network considered only 31 chemical species. 
Our results for the benchmark PDR models agreed well 
with other PDR codes considering the same 31 chemical species. 
However, the default CLOUDY chemical network contains several other chemical species, and 
we always aim 
to update reaction rates regularly \citep{{2022ApJ...934...53S},{2023RNAAS...7...45S}}. 
We use chemical reaction rate coefficients from various sources, mostly UDfA. 
Here, we present our current update, which prevents the gas-phase reaction coefficients 
from becoming large at a very high temperature.

\section{Calculations and Results}
All numerical calculations presented here are performed using the development 
version of the spectral simulation code CLOUDY, 
last described by \citet{{2017RMxAA..53..385F},{2022ApJ...934...53S},{2023RNAAS...7...45S}}. 

UDfA has been updated  over the years (RATE95 to RATE12), and in some cases, 
reactions with the problematic positive
$\beta$ have been updated to $\beta \le 0$.
Hence, we address the above-mentioned problem of rate coefficients at very high temperatures in two steps. Firstly, we update all the rate coefficients in our chemical network 
with a positive $\beta$ to RATE12 \citep{2013A&A...550A..36M}. 
There are 21 reactions for which $\beta$ has been updated to 
either a negative value or 0, 
so currently these do not diverge as the temperature increases. 
Among these, O + C$_2$ $\rightarrow$ CO + C  and 
C$_2$ + S $\rightarrow$ CS + C are important. For the first reaction, 
$\beta$ changes from 0.5 to -0.12 (RATE95 to RATE2012). For the second reaction, $\beta$ changes 
from 0.5 to 0 (RATE99 to RATE2012).
These two updates change some of the CO line intensities 
(650.074$\mu$m, 520.089$\mu$m, 433.438$\mu$m, 371.549$\mu$m, 325.137$\mu$m) 
by more than a factor of two for some standard 
PDR models (the input scripts of these models are publicly available with the \textsc{Cloudy} 
download under the directory \texttt{tsuite}). It is to be noted that C$_2$ was not a part of 
the Leiden benchmark chemical network.

The permissible upper limit of the temperature range for the UDfA reactions varies over a wide 
range: from 100 K to 41000K. 
\begin{figure}[ht]
\plotone{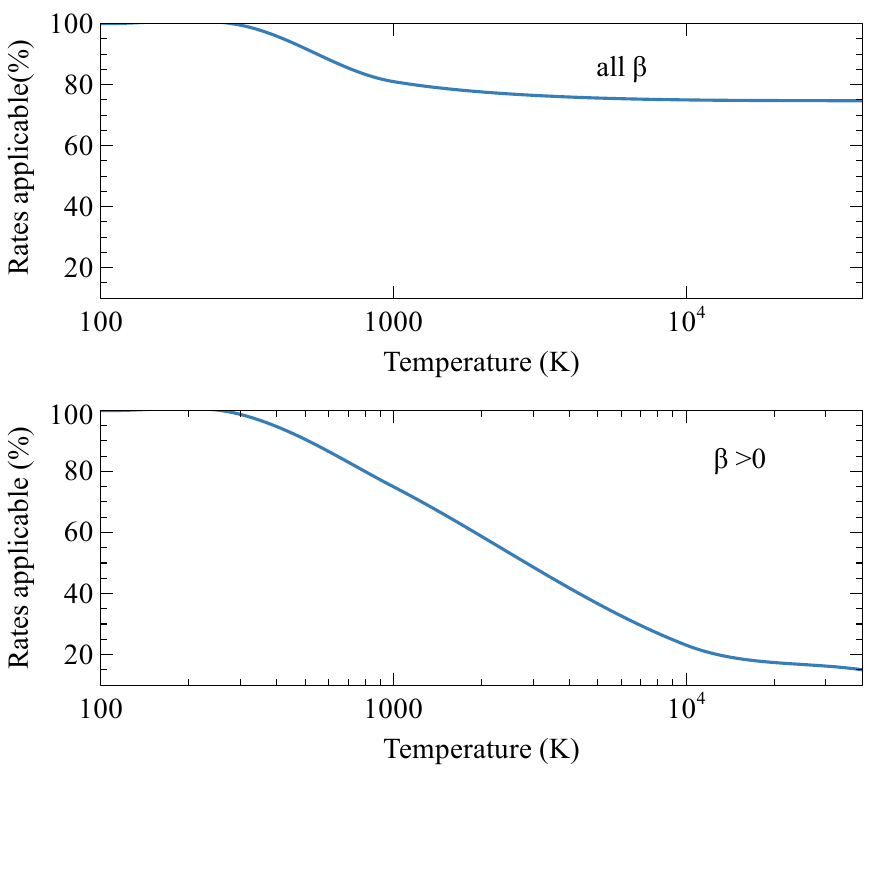}
\caption{The upper panel shows the percentage of reaction rates 
applicable for the different upper limits of temperatures (provided by UDfA) for all the UDfA rate coefficients. 
Whereas, the lower panel shows the same for $\beta$ $>0$.
\label{fig:fig1}}
\end{figure}
The upper panel of Figure \ref{fig:fig1} shows the percentage of all the UDfA rate coefficients,
for all values of $\beta$,
which are valid at that temperature. 
The lower panel shows the same for $\beta$ $>0$, the cases where extrapolation to
high temperatures can lead to unphysically large rate coefficients. 

We see that 99.5\% of all reactions are within UDfA's stated range
of validity for $T\sim 300$~K,
while 74.7\% are within range for $T \sim 4\times 10^4$~K.
Fits with $\beta > 0$ diverge at high temperatures since $k\sim T^\beta$.
98.8\% of the fits with $\beta > 0$  are within range at 300~K.
At 10$^4$~K, only 15\% of reactions with $\beta > 0$ are within range.
Unphysically large rate coefficients are predicted at high temperatures 
if the rates outside the UDfA's range are extrapolated.
These result in incorrect predictions.

We apply a temperature cap T$_{cap}$ for $\beta$ $>$0 to avoid this. 
For $T > T_{cap}$, the rate coefficients retain the same values as 
at T$_{cap}$. Though ad-hoc, we choose T$_{cap}$ = 2500K.

CLOUDY has an extensive set of test models.
The changes discussed here have little effect on most models
of molecular clouds or PDRs because the temperatures
are low enough to be within the UDfA's range of validity. 
In other sims the changes were larger, but the species had a low enough abundance 
for the effects to be unobserved.
There are large effects in warm, 5000K - 10000K, 
collisionally ionized clouds, coll\_t4 and coll\_t4\_Z30,
where the CO column 
density changed by more than 1 dex.

We hope this study motivates an extensive investigation of the high-temperature
extrapolations of the UDfA network, similar to \citet{2011A&A...530A...9R}.

\begin{acknowledgments}
GS acknowledges WOS-A grant from the Department of Science and Technology (SR/WOS-A/PM-2/2021).
GJF acknowledges support by NSF (1816537, 1910687), NASA (ATP 17-ATP17-0141, 19-ATP19-0188), and STScI (HST-AR- 15018 and HST-GO-16196.003-A).
MC acknowledges support by NSF (1910687), and NASA (19-ATP19-0188, 22-ADAP22-0139).
\end{acknowledgments}

%





\bibliography{UMIST_limitT}{}
\bibliographystyle{aasjournal}



\end{document}